# Listening to the quantum vacuum: a perspective on the dynamical Casimir effect

Gheorghe Sorin Paraoanu

QTF Centre of Excellence, Department of Applied Physics,
Aalto University School of Science, P.O. Box 15100, FI-00076 AALTO, Finland

Göran Johansson

Department of Microtechnology and Nanoscience (MC2), Chalmers University of Technology,
SE-412 96 Gothenburg, Sweden

## Abstract

Modern quantum field theory has offered us a very intriguing picture of empty space. The vacuum state is no longer an inert, motionless state. We are instead dealing with an entity teeming with fluctuations that continuously produce virtual particles popping in and out of existence. The dynamical Casimir effect is a paradigmatic phenomenon, whereby these particles are converted into real particles (photons) by changing the boundary conditions of the field. It was predicted 50 years ago by Gerald T. Moore and it took more than 40 years until the first experimental verification.

## Introduction

This year, we celebrate 50 years since the publication of the seminal paper of Gerald T. Moore [1]. This work offered a first glimpse into a puzzling quantum-field phenomenon - predicting what happens as we change the boundary conditions of an empty electromagnetic cavity, *e.g.* by moving one of its mirrors. Classically, nothing should happen - we act, in some sense, on a non-existing object.

In quantum physics, there is a time-energy uncertainty relation $\Delta E \Delta t \geq \hbar/2$ suggesting that if we consider small time-intervals $\Delta t$, we also need to consider an uncertainty of the energy of at least $\Delta E \geq \hbar/2\Delta t$. Thus, even though the vacuum has zero energy, we need to take into account the possibility of a particle with energy $\Delta E/2$ spontaneously appearing, together with its own antiparticle, and then annihilating each other again within a time $\Delta t$. There is no way we can extract this so-called zero-point energy from the vacuum, so how can we verify this very nontrivial description of nothing?

In 1970, Moore told us that if we move a mirror fast enough, we can prevent the annihilation and the particles are forced into existence. This process is called the dynamical Casimir effect (DCE). The energy is taken from the motion of the mirror and the particles should typically be created in pairs. Could this effect be observed experimentally?

# Radio signals from the vacuum

To figure this out, we need a quantitative description of this phenomenon. One interesting way to derive the result starts from a 1D gas of photons in thermal equilibrium at temperature *T*. Physically, this could be a single mode optical fiber or a microwave transmission line terminated by a black body at temperature *T*. The number of modes *dn* per unit length is *dn/L = dv/c* where ν is the frequency and *c* is the speed of light. If we assume thermal equilibrium, using the Planck average energy per mode *hv/[exp(hv/k<sub>B</sub>T) − 1]* we get the spectral energy density

$$u(\nu)d\nu = \frac{h\nu}{\exp(h\nu/k_\mathrm{B}T) - 1}\frac{d\nu}{c}.$$

The spectrum of the power emitted in one direction can be calculated by imagining that the energy *udl* contained in a certain lenght is emitted in one direction in a time *dt*, with photons moving at velocity *c = dl/dt*; therefore we get P(ν) = cu(ν). Integrating over frequency we obtain the total power emitted

$$P = cu = \frac{\pi k_\mathrm{B}^2 T^2}{12\hbar},$$

known as the Stefan-Boltzmann law in one dimension.

But how can we connect temperature to motion? First, we should recall that the vacuum is Lorentz invariant, and so we do not expect any photon emission from a mirror moving uniformly in free space. However, for an accelerating mirror such a connection exists. In the mid-1970 Paul Davies [2] and Bill Unruh [3] showed that an observer moving through vacuum with a constant acceleration *a* experiences a field at thermal equilibrium with the temperature

$$T = \frac{\hbar}{2\pi k_\mathrm{B}}\left(\frac{a}{c}\right),$$

where *c* is the speed of light. To get a measurable temperature in a laboratory, we would need to accelerate objects. Some of the highest accelerations can be obtained in the lab by using a coilgun (Gauss rifle), producing a ≈ $10^9$ m/s$^2$ with mg-mass objects. We can attach a small mirror to the bullet and ... let it shine! Unfortunately, using the equations above, we get a dismal T = 5 × $10^{-12}$ K. The emitted power corresponds to a single quanta of frequency 0.5 Hz, emitted every minute. Needless to say, this is well below any realistic experimental detection sensitivity.

But there is another way. We note that what really matters is that the electromagnetic modes get squashed by the motion of the mirror. This we can achieve, for example if the mirror does not physically move, but instead quickly changes its index of refraction.

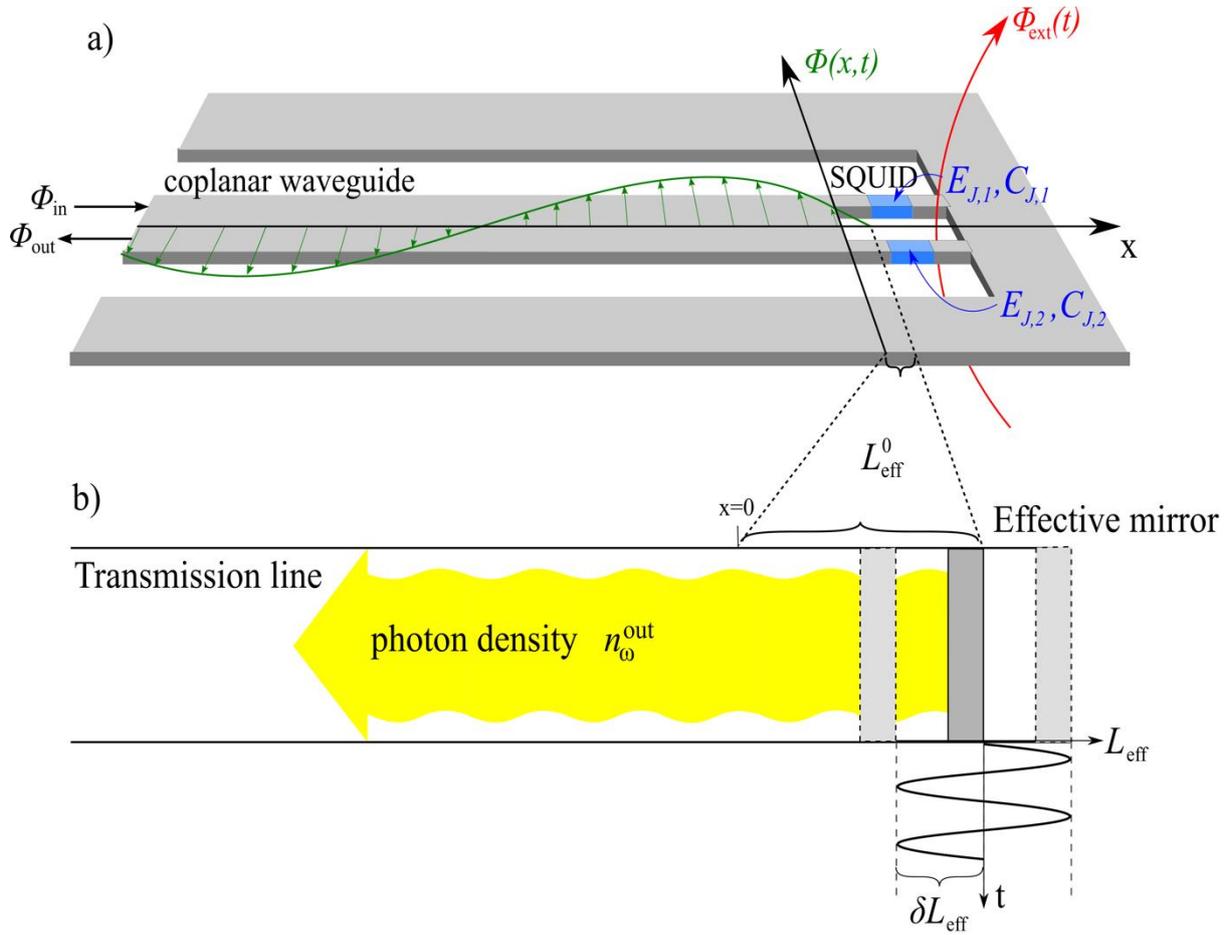

Fig. 1: a) An on-chip microwave transmission line (TL), terminated by a Superconducting Quantum Interference Device (SQUID), implementing a tunable inductance connecting the center conductor to the surrounding ground plane. b) By modulating the magnetic field through the SQUID, the electromagnetic field in the TL sees a similar boundary condition as from a moving mirror. Adapted from Ref. [6].

Consider a microwave transmission line terminated to ground through a single Superconducting Quantum Interference Device (SQUID) or an array of SQUIDs, see Fig. 1. The SQUID forms a tunable inductance, and by changing the magnetic flux threading it, it can be tuned from almost a short to a highly inductive state. This will change the standing wave pattern of the electromagnetic field in front of the mirror, from a voltage node at the mirror to an anti-node. This corresponds to moving the mirror a fraction of a wavelength, which is around a centimeter at microwave frequencies. The SQUID can be operated at 10 GHz, giving effective accelerations of around $10^{17}$ m/s², yielding a temperature of 12 mK and a power level of about -130 dBm, perfectly measurable with cryogenic microwave techniques. Just by this simple observation we get 9 more orders of magnitude compared to the coil gun, pushing the power and temperature into the experimentally observable region. Indeed, the existence of this radiation has been successfully confirmed by two experiments, one using a single SQUID [4] and the other an array of SQUIDs [5] in experiments performed at Chalmers University, Sweden and Aalto University, Finland.

In the introduction we mentioned that the particles are created in pairs together with their antiparticles and that the motion of the mirror prevents their annihilation. Also, we noted that the energy needed to create the particles is taken from the motion of the mirror. In both experiments mentioned above, the mirror was modulated harmonically with a pump frequency $f_p$ in the GHz range. The dominant process creates a pair of DCE photons, whose energies add up to one pump photon. Hence the DCE spectrum should be symmetric around the pump frequency [6, 7]. The exact shape of the spectrum depends on the density of states in the transmission line. In the Aalto experiment [5], there is a well-defined resonance resulting from the fact that the SQUID array forms a low-Q cavity, and the symmetry of the DCE spectrum is clearly visible in Fig. 2.

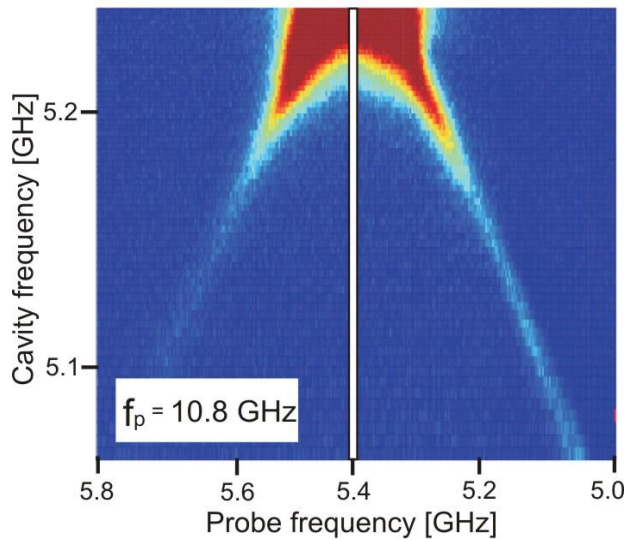

Fig. 2 Spectral power of the emitted photons recorded at various frequencies (horizontal axis). The vertical white line marks the position of half of the pump frequency. Generation of photons with an array of 250 SQUIDs, forming a cavity with relatively low quality factor of ~100. The vertical axis shows the cavity resonant frequency, which can be tuned by applying a magnetic field. The resulting sparrow-tail feature is a hallmark of the dynamical Casimir effect for this system. Adapted from Ref. [5].

**Further developments**

The pair of photons created by the dynamical Casimir effect demonstrate quantum correlations called entanglement [8]. Entanglement has been observed in both situations - in the cavity case [5] and recently in the broadband case [9]. Entanglement is a resource for quantum information processing and the dynamical Casimir effect can be used to generate this resource [10]. The entangled microwave photons could also be used as a source for a quantum radar. Photons can be generated from vacuum by using not only one, but several pumps [11], aiming at using multimode correlations as a resource [12]. And there is more to it. Vacuum can be regarded as the working fluid in quantum engines - therefore it can serve as a tool to probe the principles of thermodynamics in the quantum regime. Analog gravitational phenomena - such as the emission of Hawking radiation - have also been proposed [13].

All these show that vacuum is not inert, but instead it bursts with activity [14] that can be harnessed for quantum information processing, as well as for foundational experiments in quantum thermodynamics and analog gravity. We are willing to bet that next time you take a flight and play with those light-dimming windows, you will try to estimate how many photons you have created.

## About the authors

Sorin Paraoanu leads an experimental group focusing of superconducting circuits at Aalto University in Finland. His main research interests at present are quantum metrology and quantum simulation.

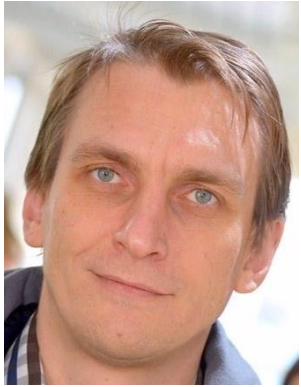

Göran Johansson is a professor in Applied Quantum Physics at Chalmers University in Sweden, leading a theory group investigating microwave quantum optics and acoustics as well as quantum algorithms.

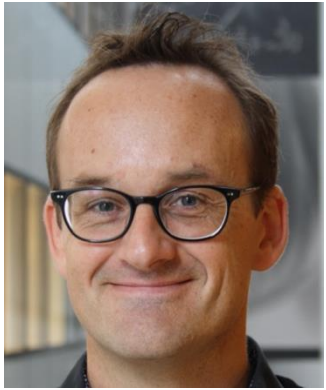

## Acknowledgements

GSP acknowledges the Academy of Finland (projects 312296,328193), EU project QUARTET (grant agreement no. 862644), and the Scientific Advisory Board for Defence of Finland.

GJ acknowledges funding from the Knut and Alice Wallenberg foundation as well as the Swedish Research Council.

Both authors would like to thank Saab for cooperation on this topic.